# A frequency-domain enhanced multi-view network for metal fatigue life prediction

Shuonan Chen, Xuhong Zhou, Yongtao Bai*

*1. School of Civil Engineering, Chongqing University, Chongqing 400045, China*

*2. Research Center of Steel Structure Engineering, Chongqing University, Chongqing 400045, China*

## Abstract

Fatigue damages and failure widely exist in engineering structures. However, predicting fatigue life for various structural materials subjected to multiaxial loading paths remains a challenging problem. A novel multi-view deep learning model incorporating frequency-domain analysis for fatigue life prediction is proposed. The model consists of two main analytical components: one for analyzing multiaxial fatigue loading paths and the other for examining the mechanical properties of materials and specimen geometrical characteristics. In the module analyzing multiaxial fatigue loading paths, convolutional neural network (CNN), long short-term memory network (LSTM), and FNet are connected in parallel to extract features individually. Features of materials and specimens are extracted through fully connected neural networks (FCNNs). Subsequently, the features from these two parts are thoroughly integrated based on attention mechanisms and connected to multiple FCNNs to accomplish fatigue life prediction. A fatigue experimental database comprising 557 samples, spanning 46 multiaxial loading paths and 19 metal materials, has been established for model training and testing. Additionally, 6 materials were respectively used as test sets to evaluate the extrapolation ability of the model. The results suggest that the proposed model exhibits robust predictive performance and extrapolation capabilities. We anticipate that the multi-view approach, along with its accuracy and applicability, can provide an unparalleled alternative for researchers in the field of engineering fatigue and beyond.

## Introduction

Fatigue failure is a significant form of damage in engineering. In sectors such as aviation, automotive, machinery, and construction, many structural components endure multiaxial cyclic loads. Fractures induced by fatigue in these components often lead to substantial damage to the overall integrity of structures. Therefore, accurate fatigue life prediction for various materials under different loading conditions is crucial.

Most fatigue life prediction models established by researchers in the past are based on summarizing experiments and built upon certain assumed conditions. These models are often empirical or semi-empirical, and their accuracy is frequently limited due to researchers' inability to comprehend the intricate physical phenomena. In recent years, research utilizing machine learning methods for fatigue life prediction has emerged as a new trend. This approach does not rely on any assumptions and can bypass various constitutive relationships implicit in physical phenomena, solely based on data for modeling purposes.

The application of machine learning algorithms in fatigue life prediction has been widely studied. For the prediction of uniaxial fatigue life, Wu and Bao et al. explored the potential factors affecting the fatigue life of selective laser melted Ti-6Al-4 V alloy using support vector machine(SVM)[1]; He et al. used Random Forest to predict the fatigue life of AISI 4140 and CA6NM, and the data was mainly concentrated within a 2 times scatter band[2]; Zhang and Sun et al. used the neuro-fuzzy-based machine learning method to predict the high cycle fatigue life of laser powder bed fusion welded stainless steel 316L[3]; Srinivasan et al. evaluated the fatigue life of 316L (N) stainless steel under creep conditions using ordinary artificial neural network (ANN)[4].

For multiaxial fatigue loading, Krzysztof PałCzynski et al. predicted the fatigue life of PA38-T6 aluminum alloy using ANN[5]; Yang et al. effectively extracted the temporal features of the loading path by vectorizing it and inputting it as temporal data into LSTM, and the prediction results mostly fell within the 1.5 times error scatter band[6]; In addition, Yang et al. also applied self-attention mechanism to characterize the influence of complex loading history and temperature changes on fatigue life, showing good ability in predicting multiaxial fatigue life in mechanical and thermo mechanical fields[7]; Sun et al. proposed a CNN hysteresis line image recognition model for multiaxial low cycle fatigue life prediction by using loading paths as image inputs to extract features from CNN[8]; Heng et al. used CNN-LSTM to enhance the extrapolation ability of the model under different loading paths[9]; Chen et al. established a physics-informed neural network (PINN) and achieved good prediction accuracy based on a small dataset[10]; Zhou et al. established an ANN model with knowledge based features to predict the multiaxial low cycle fatigue life under irregular loads and validated it on 304L stainless steel[11]. Wang et.al studied the influence of defect characteristics on fatigue life by combining a Gaussian distribution-based data augmentation technique with PINN[12].

The aforementioned research indicates that deep learning demonstrates stronger applicability in multiaxial fatigue life prediction. Deep learning possesses powerful feature extraction capabilities, enabling it to effectively learn deep patterns and representations within multiaxial fatigue loading data[13]. However, establishing a model suitable for multiple metal materials that can identify complex loading paths still remains challenging.

Currently, proposed deep learning models usually analyze loading paths from a single perspective. Moreover, due to the exorbitant cost of fatigue testing, the available training data are limited. Inputting limited fatigue loading data into a single network structure for feature extraction may result in the network inadequately capturing deep features at different levels of complexity within the loading paths, or even losing partial information from the input data. Consequently, this can lead to decreased generalization ability of the model and an increased likelihood of overfitting. For instance, inputting data into a CNN framework primarily extracts features at the convolution level, while inputting data into an LSTM framework extracts features from the time domain. Thus, models trained in this manner often exhibit poor extrapolation capabilities[6, 14], with performance deteriorating when tested on complex loading path datasets after being trained on simpler loading path datasets. Additionally, current deep learning models that can recognize complex loading paths are trained and established based on datasets of specific types of materials. Consequently, the trained models are only applicable to the corresponding material datasets. Predicting the fatigue life of other types of materials requires additional training on datasets specific to those materials, lacking a universal model applicable to multiple materials simultaneously.

Fourier transform can convert the original time-domain signals into frequency-domain representations, thus revealing many features that are difficult to capture in the time domain. In neural networks, the Fourier transform has found widespread application. For instance, the application of the fast Fourier transform in CNNs can significantly enhance training speed without sacrificing accuracy[15]–[17], while the discrete Fourier transform can be used in RNNs to effectively address the issues of vanishing and exploding gradients[18, 19]. Fnet, proposed by James et al., is a variant of the Transformer architecture that replaces self-attention layers with Fourier layers[20]. Through Fourier transformation, Fnet facilitates interaction between internal and external information of tokens, enabling comprehensive integration of token information. This study suggests that Fnet is an efficient hybrid mechanism.

This study aims to introduce frequency-domain analysis and conduct multi-level, multi-angle feature extraction on fatigue data. Subsequently, through ensemble modeling, the study integrates Fnet, LSTM, and CNN networks to construct a multi-view neural network model. On this basis, to solve the inability to apply on multiple materials simultaneously, the model is divided into two modules for feature extraction: one for analyzing fatigue loading paths and the other for material characteristics. These two sets of features are then thoroughly concatenated and integrated through a self-attention mechanism, ultimately predicting material fatigue life using FCNN. The fatigue loading path analysis module comprises three neural networks in parallel: CNN for extracting spatial features of the data, LSTM for capturing time-dependent relationships, and Fnet for capturing frequency-domain features. The analysis of input data is presented in Section 2. The specific structure of the proposed model is detailed in Section 3. In Section 4, fatigue data from 19 materials[21]–[35] under multiaxial loading conditions are collected to analyze the capabilities of the proposed model. The data are divided into two parts: one consisting of multiaxial fatigue loading path data, and the other consisting of mechanical properties and geometric characteristics of corresponding standard specimens for each material. Additionally, 6 materials containing representative complex loading paths are selected to test the extrapolation capabilities of the proposed model. Finally, Section 5 summarizes the conclusions.

## 2 Multiaxial fatigue experiments and dataset establishment

### *2.1 Multiaxial fatigue experiments*

There are various shapes of multiaxial fatigue test specimens, including plate-shaped, disk-shaped, cross-shaped, and thin-walled tubular specimens. Among them, thin-walled tubular specimens are the most commonly used. This type of specimen can be subjected to various forms of loading, such as tension, torsion, internal pressure, and external pressure, to obtain various desired stress or strain states. Additionally, the stress state of thin-walled tubular specimens is straightforward, and stress can be directly calculated from the applied loads. Furthermore, to reduce the gradient of shear strain along the thickness direction of the tube wall, the tube wall is typically made very thin, approximately 1-2 mm.

Multiaxial cyclic loading methods include torsion-bending compound loading, tension-compression-torsion compound loading, biaxial loading of cross-shaped specimens, tension-compression-torsion compound loading of thin-walled tube specimens, and tension-compression loading, among many others. Among these loading methods, tension-compression-torsion compound loading is currently widely adopted[36, 37]. A typical multiaxial tension-compression fatigue specimen is illustrated in Fig. 1, with the central portion designated as the measurement section, and the thicker portions at both ends subjected to force or deformation by the experimental equipment. The $L_g$ in Fig. 1 represents the gauge length, $d_i$ represents the inner diameter, and $d_o$ represents the outer diameter.

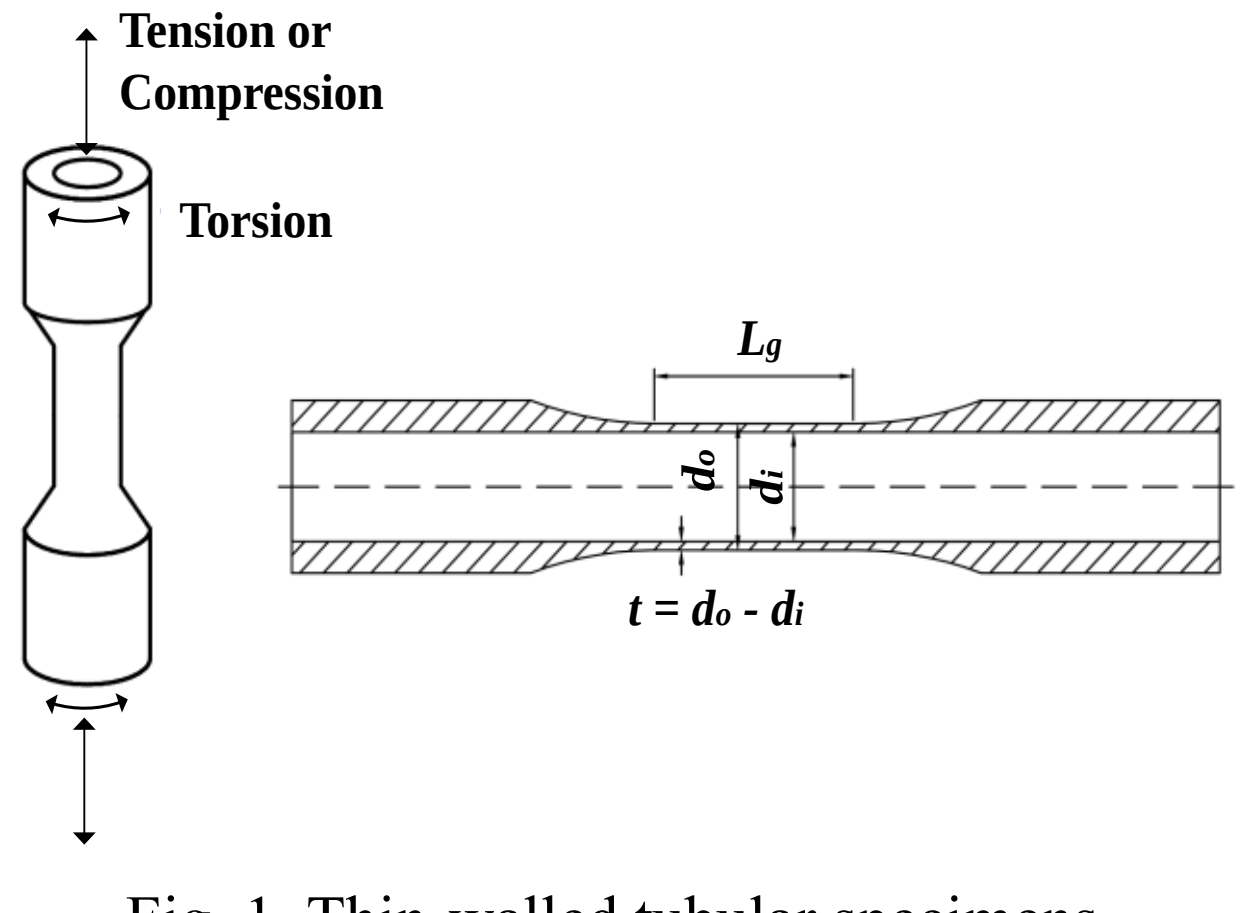

Fig. 1. Thin-walled tubular specimens.

Experiments can be categorized into stress-controlled experiments and strain-controlled experiments based on the loading process. In this study, we primarily focus on strain-controlled multiaxial fatigue loading tests. Under different waveforms of experimental loading paths, the loading paths vary significantly[23, 24, 32, 35]. To enable the model to accurately distinguish between the intricate and diverse loading paths, it is crucial to collect a sufficient number of samples to train the model effectively.

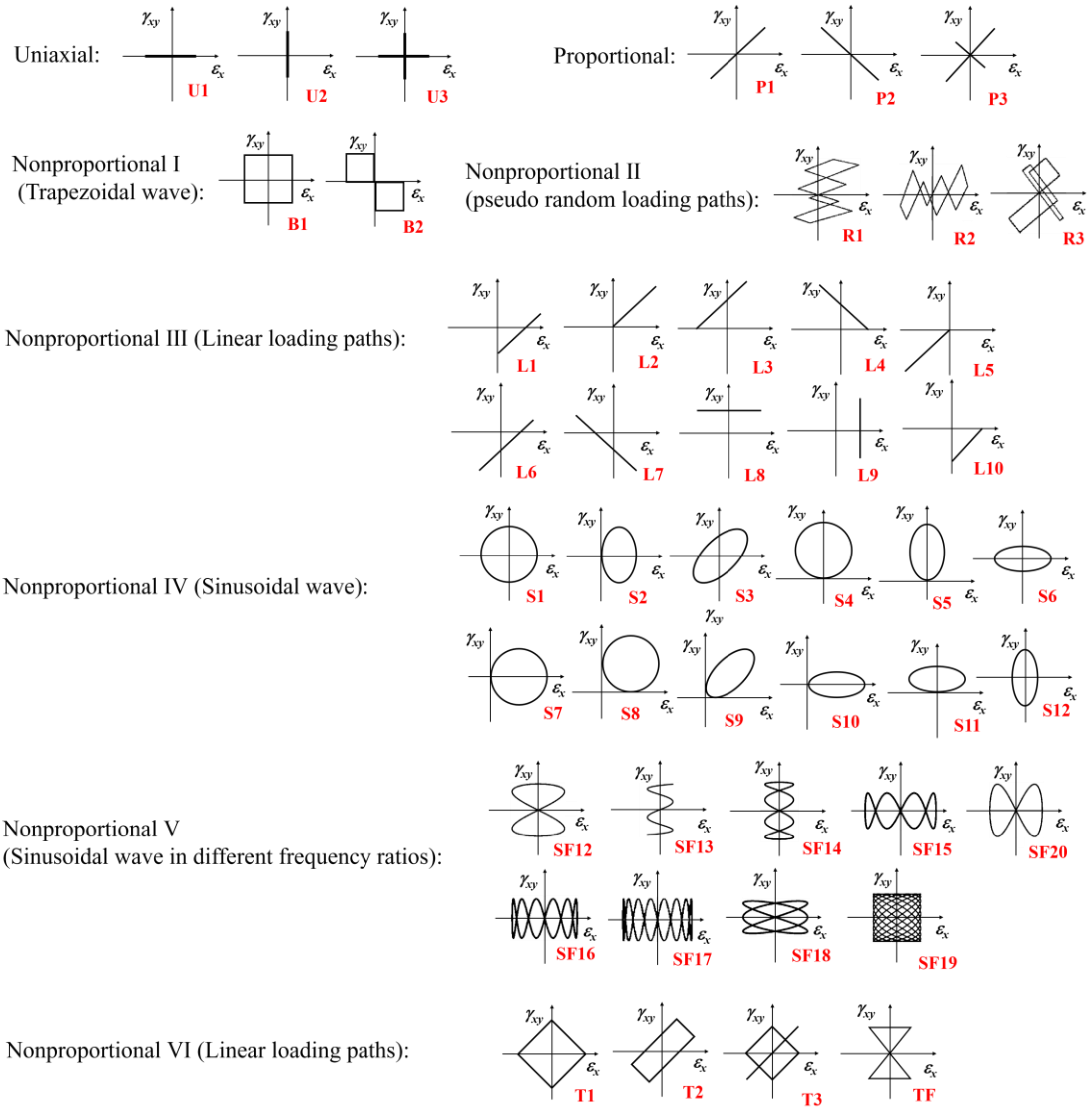

Fig. 2. Various loading paths collected from 19 different metal materials.

In this study, 557 samples of tension-torsion fatigue loading data are collected from 19 different materials, including carbon steel, alloy steel, aluminum alloy, titanium alloy, and others. The data encompass various loading scenarios including categories, such as uniaxial loading, multiaxial proportional loading, multiaxial nonproportional loading, and pseudo-random loading, as illustrated in Fig. 2. The mechanical properties obtained from the experiments for these materials, as well as the geometric dimensions of the middle gauge section of the designed specimens, are summarized in Table 1.

Among them, the sinusoidal loading paths also include multiple categories that consider normal mean strain, shear mean strain, normal shear and normal strain, and different loading frequency ratios of tension and torsion. This provides a foundation for the model to recognize features of different loading paths during the training process.

### *2.2 Data analysis*

The multiaxial fatigue life of materials can be summarized as being influenced by three main aspects: the material aspect, the loading aspect, and the environmental aspect[38]. Firstly, at the material aspect, microstructure, processing methods, mechanical properties, and geometric dimensions play crucial roles in fatigue life. Secondly, loading applied to the material is also an important influencing factor, specifically manifested as loading paths in mult-axial fatigue. Thirdly, factors such as temperature and humidity in the external environment also have an impact on fatigue life. Accurately predicting the multiaxial fatigue life of different materials requires considering various complex influencing factors. Establishing a universal neural network model applicable to different materials and loading paths entails establishing the correlation between these influencing factors and fatigue life. This requires the comprehensive selection of appropriate features from various factors affecting fatigue life as inputs to the neural network.

Table 1. 19 specific material parameters and corresponding loading paths

| Materials | Loading Paths | $E$(GPa) | $\sigma_y$(MPa) | $\sigma_u$(MPa) | $\nu$ | $d_i$(mm) | $d_o$(mm) | $L_g$(mm) | Samples |
|---|---|---|---|---|---|---|---|---|---|
| Q235b[21] | U1,U2,P1,S1, S3,T1,T2 | 206 | 235 | 412 | 0.304 | 14.5 | 16.5 | 50 | 38 |
| HRB335[22] | U1,U2,S1, T1,TF | 210 | 355 | 520 | 0.300 | 14.5 | 16.5 | 50 | 23 |
| E235[23] | U1,U2,P1,S1, | 196.4 | 247.8 | 375.4 | 0.300 | 8 | 11 | 26 | 55 |
| | SF12,SF14,SF15,SF18 | | | | | | | | |
| E355[23] | U1,U2,P1,S1,SF12,SF14,SF15,SF16,SF19 | 208.6 | 318.4 | 473 | 0.290 | 8 | 11 | 26 | 59 |
| X5CrNi18-10[23] | U1,U2,P1,S1,SF12,SF13,SF15,SF16,SF19 | 200.8 | 265 | 654.4 | 0.290 | 7 | 9.7 | 25 | 63 |
| SA 333 Gr.6 [24] | U1,U2,P1,S1,B1,T1,T2 | 203 | 307 | 463 | 0.300 | 22 | 25.4 | 30 | 39 |
| 45#[25] | U1,U2,L2,S1,S2,S3,S4,S5,S6,S7,S8,S9,S10,S11,S12 | 206 | 370 | 610 | 0.300 | 21 | 25 | 50 | 21 |
| AISI 316L[26] | U1,U2,P1,S1,S3 | 193 | 272 | 610 | 0.300 | 7 | 10 | 30 | 25 |
| S460N[28] | U3,P1,S1,S3,S7,B1,T1,SF12,SF20,L1,L3,L8,L9, L10 | 208.5 | 500 | 643 | 0.300 | 36 | 41 | 2 | 39 |
| 16MnR[29] | S11,P1,L6,L7 | 212.5 | 324.4 | 544.5 | 0.310 | 20 | 24 | 29.9 | 15 |
| 1045HR[27] | P1,S1,S6,S12 | 205 | 382 | 652 | 0.290 | 25 | 29 | 33 | 28 |
| Mild[30] | P1,S1,S3 | 210 | 210 | 316 | 0.404 | 16.5 | 19.5 | 12 | 10 |
| 304 stainless steel[31] | U1,U2,P1,S1,B1,B2 | 183 | 325 | 650 | 0.105 | 25.4 | 33 | 25.4 | 8 |
| Inconel 718[32] | P1,L1,L2,L3,L4,L5 | 209 | 1160 | 1445.31 | 0.300 | 25 | 29 | 25 | 12 |
| PA38-T6[23] | U1,U2,P1,S1,SF12,SF13,SF15,SF17,SF19 | 68.3 | 191.5 | 229.1 | 0.350 | 10 | 13 | 26 | 64 |
| 7075-T651[33] | U1,P1,S1,SF12,SF14 | 71.7 | 501 | 561 | 0.306 | 20 | 24 | 29.9 | 19 |
| Pure Ti[34] | U1,U2,P1,S1,S3 | 112 | 475 | 558 | 0.400 | 10 | 11.5 | 20.1 | 12 |
| BT9[34] | U1,U2,P1,S1 | 118 | 910 | 1080 | 0.370 | 10 | 11.5 | 20.1 | 6 |
| SNCM630 [35] | U1,U3,P1,P3,S1,B1,T1,T2,T3,R1,R2,R3 | 196 | 951 | 1103 | 0.273 | 10 | 12.5 | 30 | 17 |

In this study, the samples collected are all under room temperature conditions, so we consider fixing the external environmental conditions, and the established model is only based on room temperature conditions. At the material level, four mechanical features—namely, elastic modulus (E), yield strength ($\sigma_y$), tensile strength ($\sigma_u$), and Poisson's ratio ($\nu$)—are selected, along with three geometric features of the specimens ($d_i$, $d_o$, $L_g$). At the loading level, multiple different multiaxial fatigue loading paths are considered.

Due to the diversity and complexity of loading paths, representing these paths using structured data inevitably leads to the loss of some information present in the data. To ensure the complete input of information from the loading paths into the neural network, it is necessary to transform the data into a suitable format. As depicted in Fig. 3, a circular path in multiaxial fatigue loading can be viewed as a data sequence containing multiple loading points, represented in vector form as: ($X_1$, $X_2$, $X_3$, ..., $X_{t-1}$, $X_t$). In the case of strain-controlled tension-torsion loading, this can be written as ($\varepsilon_x$; $\gamma_{xy}$), and for a more general scenario, it can be expressed as ($\varepsilon_x$; $\varepsilon_y$; $\varepsilon_z$; $\gamma_{xy}$; $\gamma_{xz}$; $\gamma_{yz}$)[6]. This sequential data can effectively encompass essential information from the loading path, such as amplitude, frequency, mean loading and changes in loading direction. Vectorizing the loading paths allows them to be treated as sequential information input into the neural network.

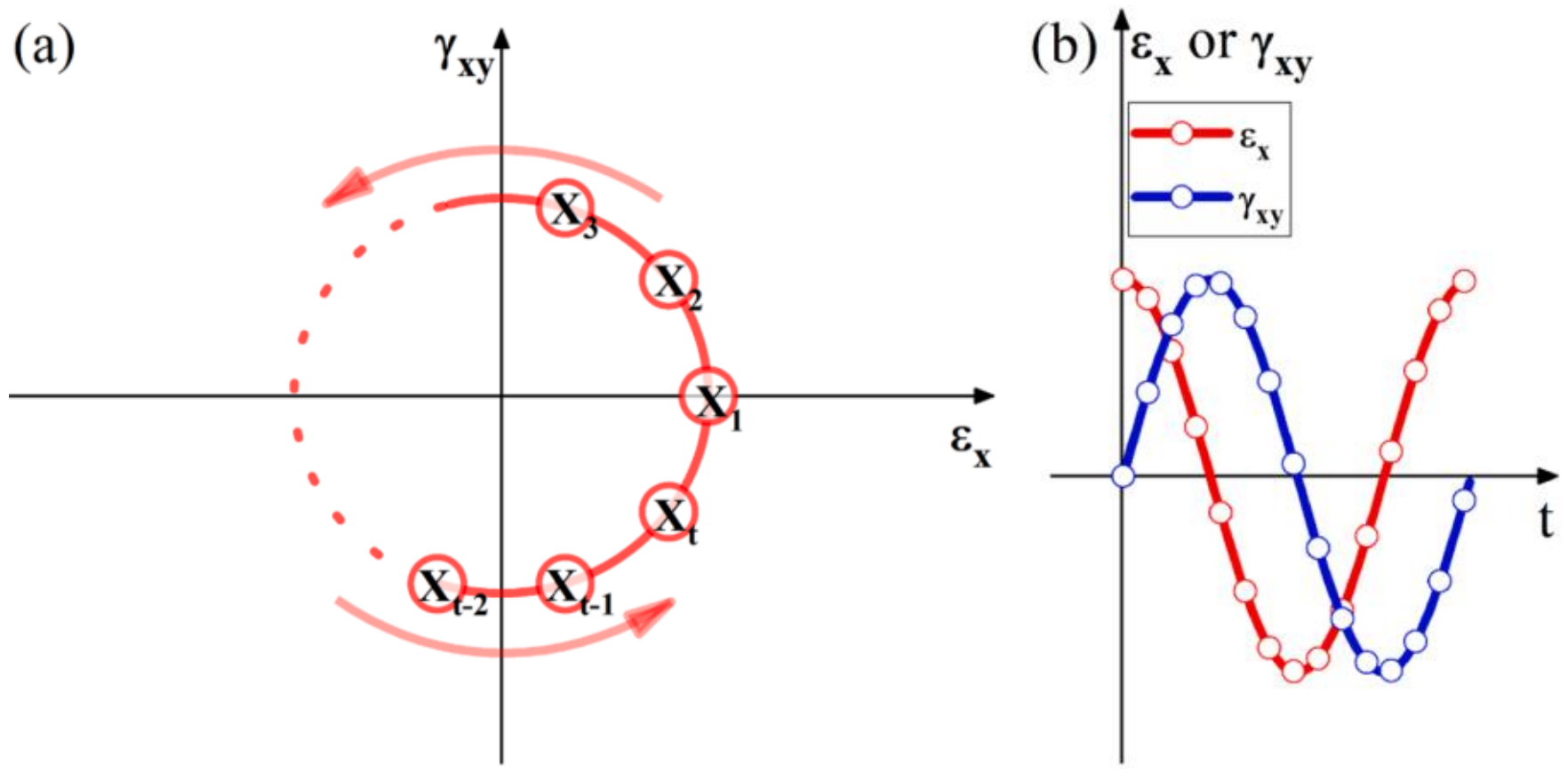

Fig. 3. Vectorized loading paths.[6]

The mechanical properties and geometric dimensions of materials are treated as structured data, while the loading path information is regarded as sequential data. These two types of data exhibit significant differences in terms of dimensions and structures, a point that will be elaborated on in Section 3.1. However, a single neural network structure is inadequate for handling both types of data. The proposed model in this study with multi-module can simultaneously consider and process both time series data and structured data. Each module can be optimized and adjusted for different types of data, enabling the model to better adapt to the input data types and enhance its generalization ability. Each module of the model can be specifically responsible for extracting features from specific types of data. Subsequently, through a self-attention mechanism, the features from these two parts are thoroughly integrated, resulting in a more comprehensive and enriched feature representation.

## 3. Framework of the proposed method

The processing flow of the model proposed in this study is divided into three stages. The first stage is to establish a dataset, which involves preprocessing the collected data and dividing it into training and testing sets. The second stage is the model establishment stage, which requires inputting the training set made in the first stage into the constructed neural network model to complete the model training. The final stage evaluates the trained model using the test set. Fig. 4 is a specific flowchart that vividly illustrates how these three stages are completed in sequence.

Section 3.1 will introduce the inputs of the model, 3.2 will provide a detailed explanation of the specific framework and composition of the model, and 3.3 will analyze and determine the hyperparameters and other parameters within the model.

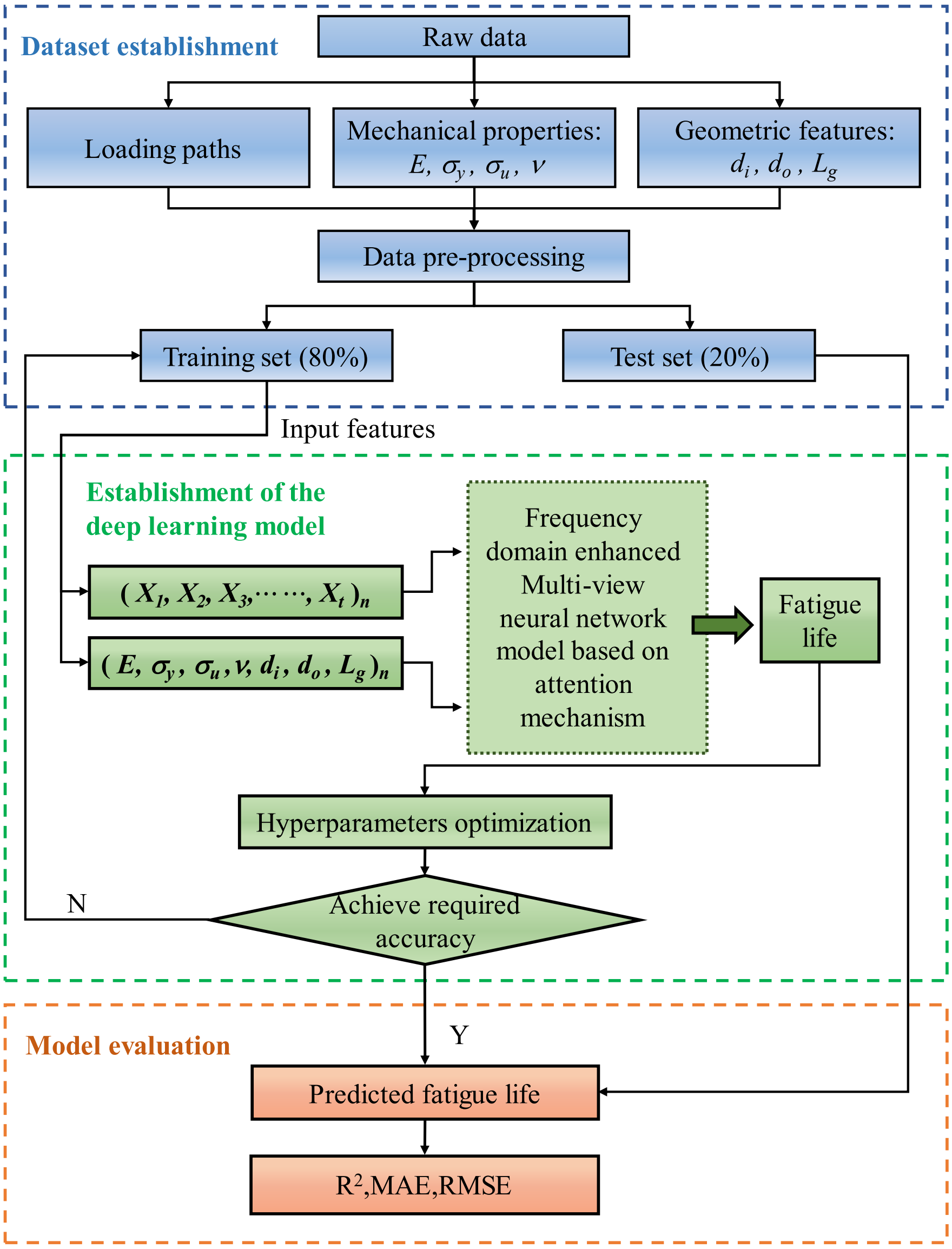

Fig. 4. Flowchart of the proposed method.

### *3.1 Input of the model*

#### *3.1.1Input form of the loading paths and material properties*

As described in Section 2.2, the model proposed in this study consists of two input components. One component comprises time-series data formed by the loading paths, while the other consists of structural data. These two data components will be input into specialized processing modules within the model, where features extracted from these two data components will generate corresponding prediction values. The vectorized loading paths can be represented as:

$$X_{Li} = \begin{bmatrix} \varepsilon_{\mathrm{x}}^{1} & \gamma_{xy}^{1} \\ \varepsilon_{\mathrm{x}}^{2} & \gamma_{xy}^{2} \\ \varepsilon_{\mathrm{x}}^{3} & \gamma_{xy}^{3} \\ \vdots & \vdots \\ \varepsilon_{\mathrm{x}}^{\mathrm{t}-1} & \gamma_{xy}^{t-1} \\ \varepsilon_{\mathrm{x}}^{\mathrm{t}} & \gamma_{xy}^{t} \end{bmatrix} \tag{1}$$

In the equation, $X_{Li}$ represents the loading path of the ith sample in the dataset, $\varepsilon_{\mathrm{x}}^{\mathrm{t}}$ denotes the axial strain at the interval within the segmented time frame, and $\gamma_{\mathrm{x}}^{\mathrm{t}}$ represents the corresponding shear strain.

Structural data can be represented as:

$$X_{Si} = [E_i \quad \sigma_{yi} \quad \sigma_{ui} \quad v_i \quad d_{ii} \quad d_{oi} \quad L_i] \tag{2}$$

where $X_{si}$ represents the material characteristic matrix of the ith sample, and $E_i$ in the matrix represents its corresponding elastic modulus, yield strength, tensile strength, Poisson's ratio, the inner diameter of the gauge section, outer diameter of the gauge section, and the gauge length.

The $n_t$ in Fig. 5 represents the length of the vectorized loading path, and the subscript $n$ represents the number of batches. The role of a neural network model can be seen as a complex function fitter, which ultimately obtains a function between $X_{Li}$, $X_{Si}$, and fatigue life Nf by continuously changing internal parameters:

$$f(X_{Li}, X_{Si}) = N_f \tag{3}$$

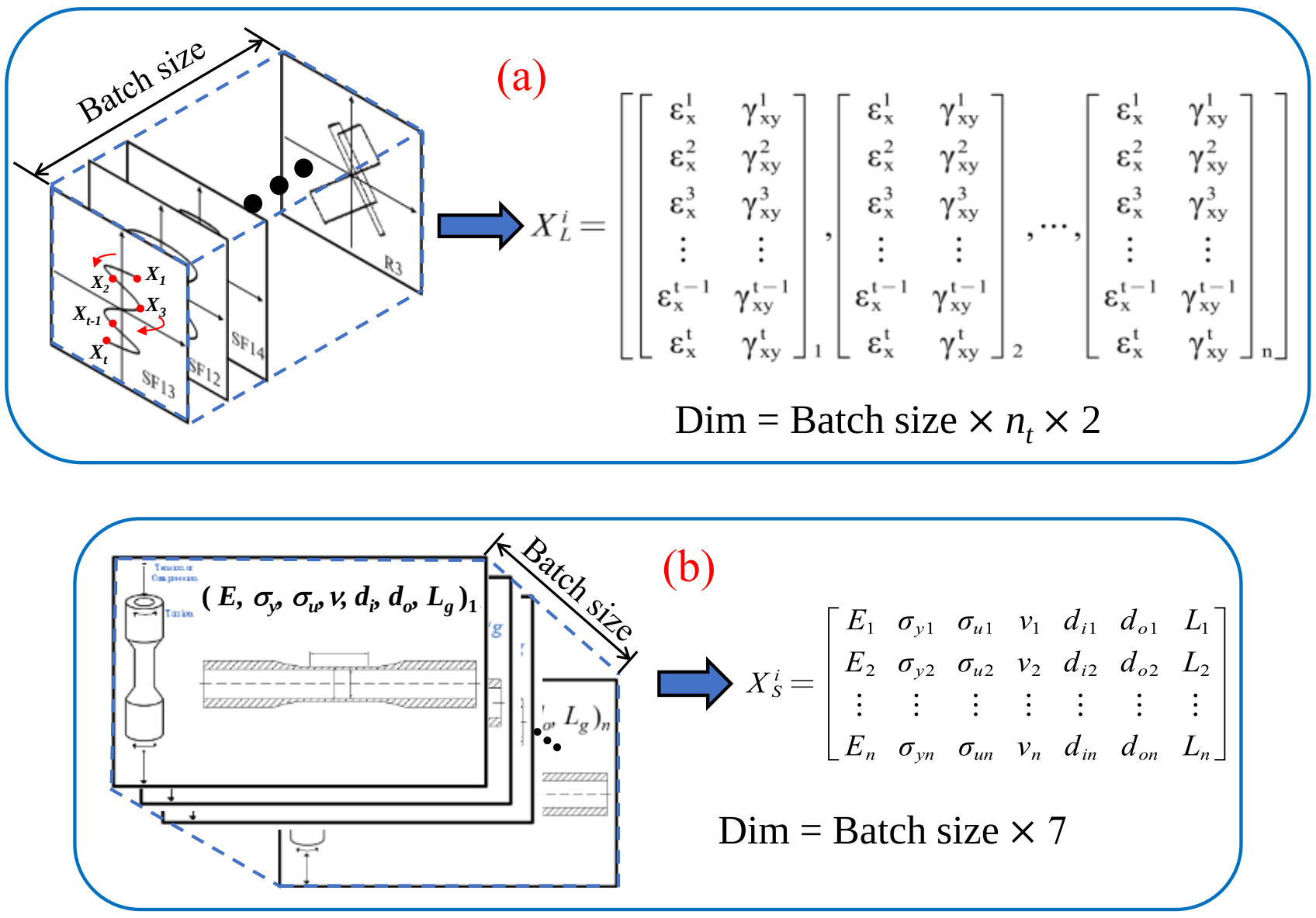

Fig. 5. (a)Input of loading paths (b)Input of geometric characteristics of material mechanical properties.

In Eq. (3), $f$ represents the complex nonlinear relationship between the input $X_{Li}$ and $X_{Si}$ constructed by the neural network model, and the output fatigue life $N_f$. It should be noted that in more general cases, if additional factors affecting fatigue life are considered, such as temperature, humidity, and other external conditions, as additional inputs, the relationship may become even more intricate.

$$X_{Ei} = [T_i, H_i, ..., E_i] \tag{4}$$

Where $T_i$ and $H_i$ respectively denote the temperature and humidity corresponding to the ith specimen during the experiment, while $E_i$ represents the values of any other external influencing factors that need to be considered. Thus, the mapping relationship represented by the neural network can be expressed in a more general form as:

$$f(X_{Li}, X_{Si}, X_{Ei}) = N_f \tag{5}$$

The data collected in this study all originate from experiments conducted under normal room temperature and humidity conditions. Therefore, the value of $X_{Ei}$ is essentially a constant and is not included as an input to the neural network. Consequently, the input for each training batch (batch size, i.e., the number of samples input to the model at each iteration) can be represented as shown in Fig. 5. The input for the loading paths is a three-dimensional tensor $X_L^i$, with a specific dimension size of batch size $\times n_t \times 2$, while the input for material properties is a two-dimensional tensor $X_S^i$, with a specific dimension size of batch size $\times 7$. Where $i$ represents the input of different batch numbers.

### *3.1.2 Data preprocessing before training*

Before formally training the model, data normalization is a common preprocessing step. Its purpose is to scale the values of different features to a similar range, ensuring that the model is more stable and converges faster during training[6, 9, 10, 11]. Specifically, normalization helps to mitigate the impact of scale differences between different features on model training, allowing the model to better learn the relationships between the data[39].

In this study, min-max scaling is employed for structured data, which involves linear transformation of the original data to map the resulting values to the range [0,1]. The transformation function is as follows:

$$x_{new} = \frac{x - x_{min}}{x_{max} - x_{min}} \tag{6}$$

where $x_{max}$ represents the maximum value of the sample data, and $x_{min}$ represents the minimum value of the sample data, which corresponds to the mechanical properties ($E$, $\sigma_y$, $\sigma_u$, $v$), as well as the geometric dimensions ($d_i$, $d_o$, $L_g$) of each material. For instance, in the normalization process of the elastic modulus $E$:

$$E_{norm} = \frac{E - E_{\min}}{E_{\max} - E_{\min}} \tag{7}$$

where $E_{\max}$ represents the maximum elastic modulus among the 19 materials, $E_{\min}$ represents the minimum value, and $E_{norm}$ is the normalized value. The same normalization process is applied to the other 6 values.

Due to the wide range of fatigue life $N_f$ for materials (spanning from hundreds to millions), it is common to take the logarithm of the fatigue life $N_f$, denoted as $\lg N_f$, for prediction. This approach ensures a more uniform and stable distribution of data, thereby enhancing the training effectiveness and predictive performance of the model.

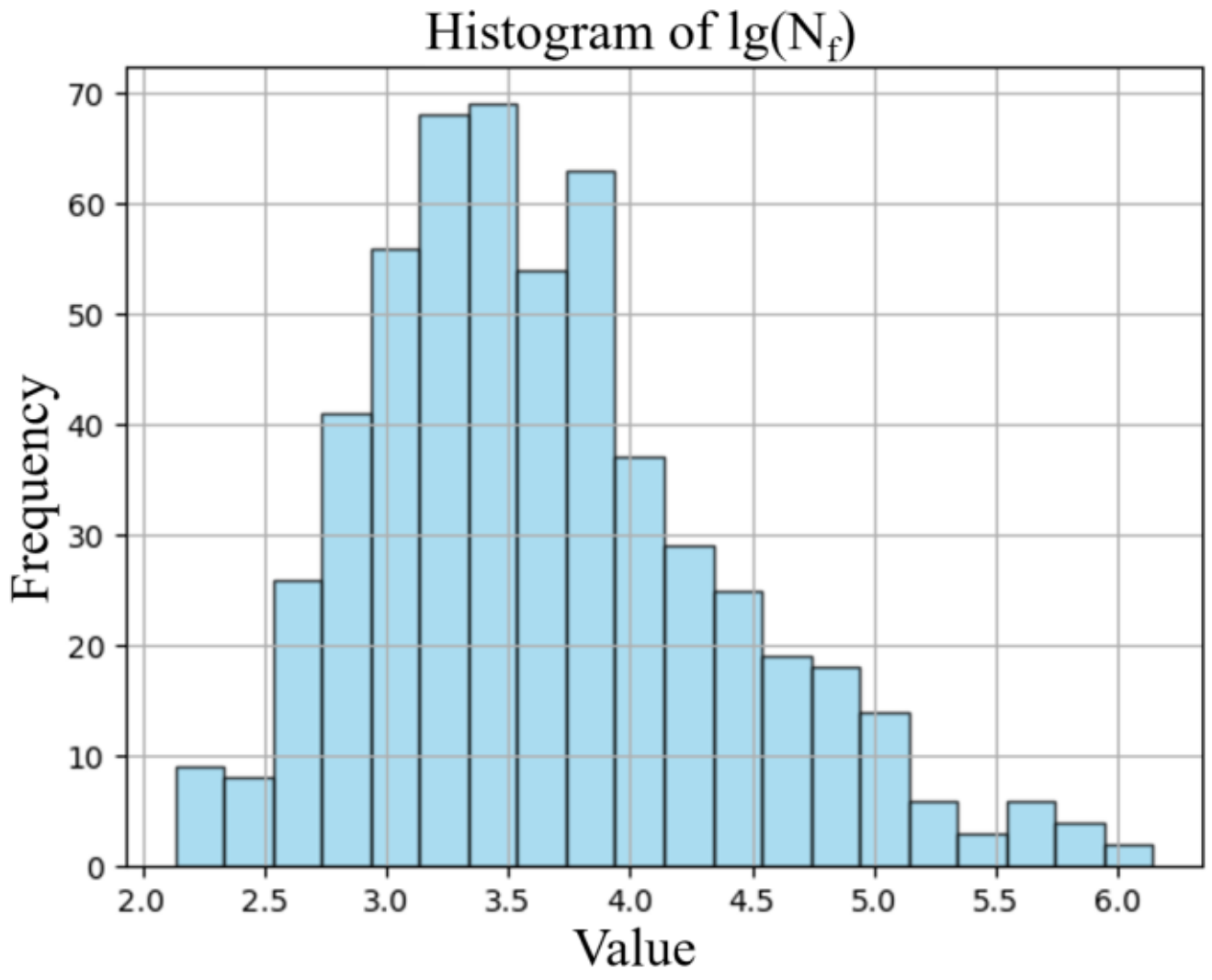

Fig. 6. Distribution of logarithmic lg ($N_f$) values of fatigue life of collected samples

Fig.6 shows the distribution of fatigue life for 557 samples collected in this study, with the highest values between 3.0-4.0 (corresponding to fatigue life 1000-10000). The logarithmic distribution of multiaxial fatigue life for approximately 250 specimens is in this range, accounting for approximately 45% of the total sample size; The minimum value is around 2.2 (corresponding to a fatigue life of about 150 cycles), and the maximum value is around 6.2 (corresponding to a fatigue life of about 1.6 million cycles). The distribution range of fatigue life values involved in the entire dataset is wide, ranging from hundreds to millions, which effectively ensures the richness and diversity of the data.

***3.2 Structure of the proposed frequency-enhanced multi-view neural network model***

The specific architecture of the multi-view neural network model established in this paper is illustrated in Fig. 7. The model is composed of 5 main steps to predict fatigue life: first, input data; second, feature extraction from the two input components; then, feature fusion; the fused features are fed into a fully connected neural network to predict fatigue life; finally, output the results. The parameters of the model will be discussed in Section 3.3.

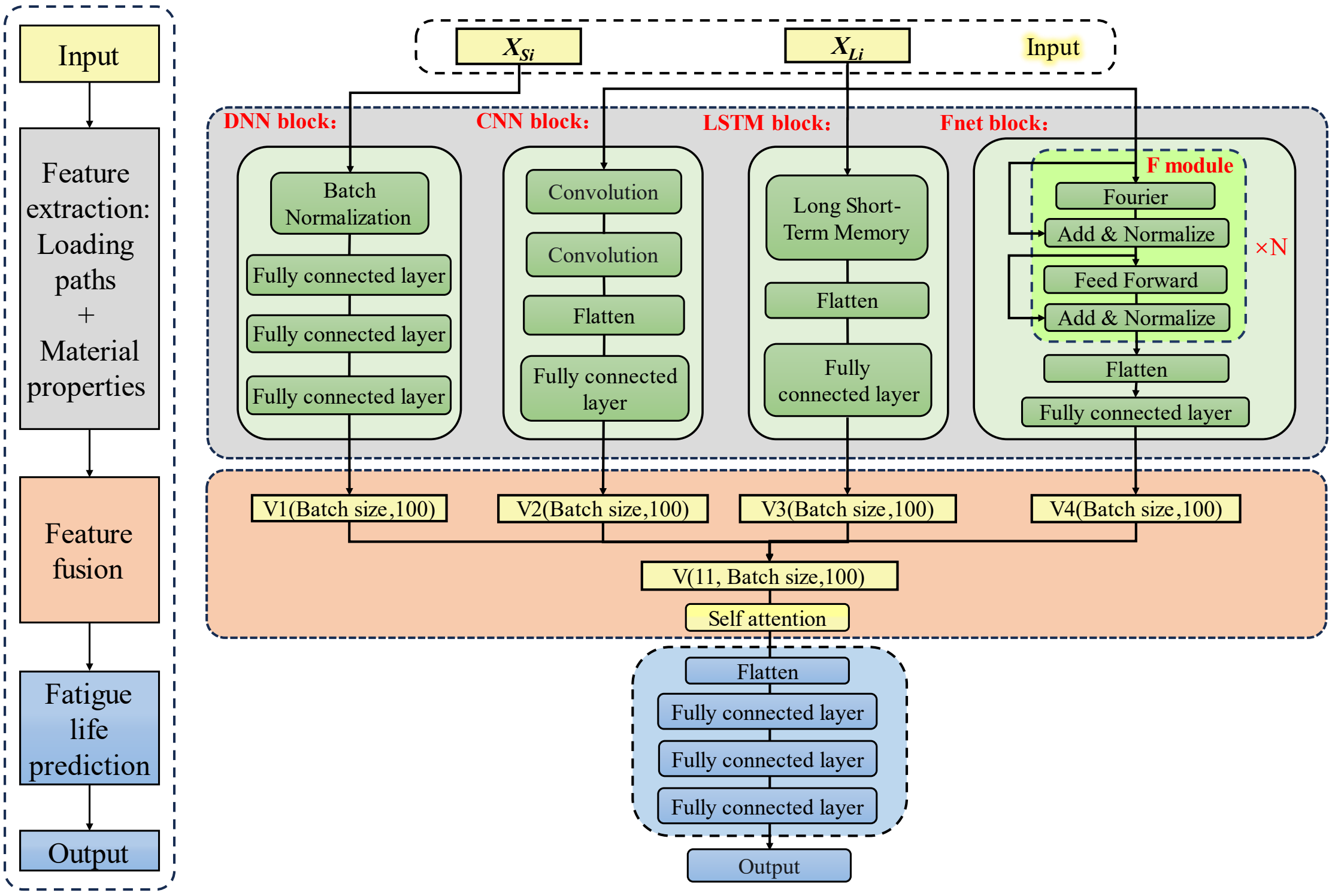

Fig. 7. 5 main steps of the proposed multi-view deep learning neural network.

### *3.2.1 Feature extraction*

Based on the aforementioned input, the structured data $X_{Si}$ is fed into the Deep Neural Network (DNN) module. This module first applies batch normalization to the input data. Batch normalization is a data normalization technique that accelerates the convergence speed during model training, making the training process more stable[40]. After completing batch normalization, the structured data is passed through three fully connected neural network layers to extract features. Finally, the module outputs a feature vector **V1** with dimensions of batch size × 100, as illustrated in Fig. 8. It should be noted that the "100" here refers to the number of neurons at the fully connected layer, which is a hyperparameter that needs to be determined.

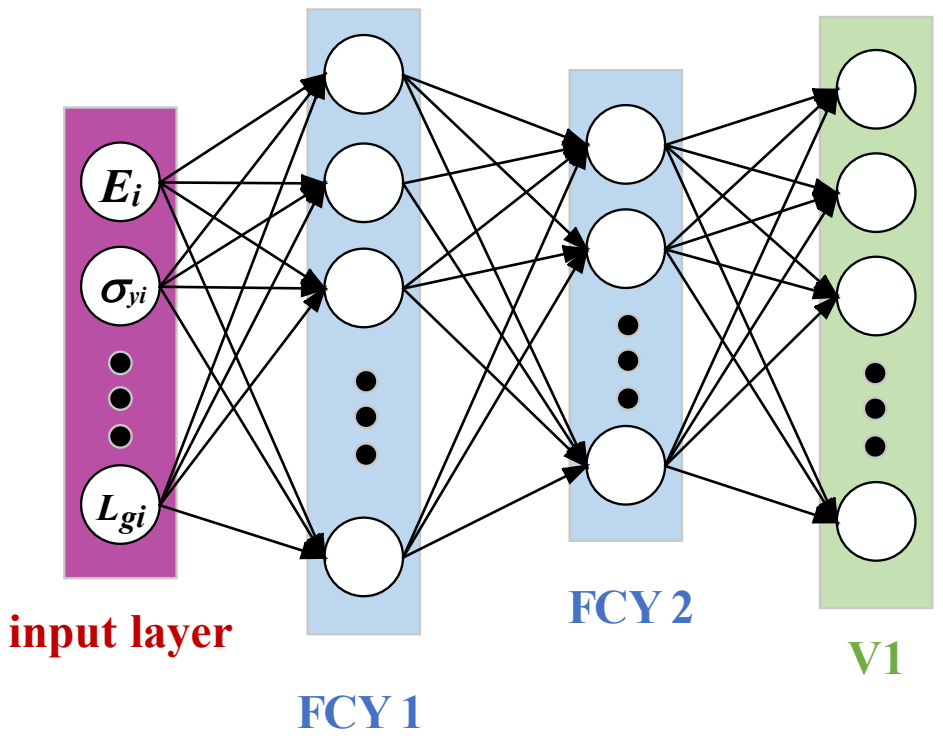

Fig. 8. DNN block feature extraction process

The multiaxial fatigue loading paths $X_{Li}$ are separately input into the constructed CNN block, LSTM block, and Fnet block. These three modules perform feature extraction from the loading paths from three different perspectives: convolution, time domain, and frequency domain, respectively.

The internal structure of the multi-view analysis module is shown in Fig. 7. The CNN module utilizes a convolutional neural network. After normalization, the data is input into the convolutional layers, then flattened into a one-dimensional vector, and finally passed through a fully connected layer[5, 6, 13] to obtain a feature vector **V2** with dimensions of batch size × 100. Fig. 9 vividly illustrates how the data is convolved and then output V1 through a fully connected layer.

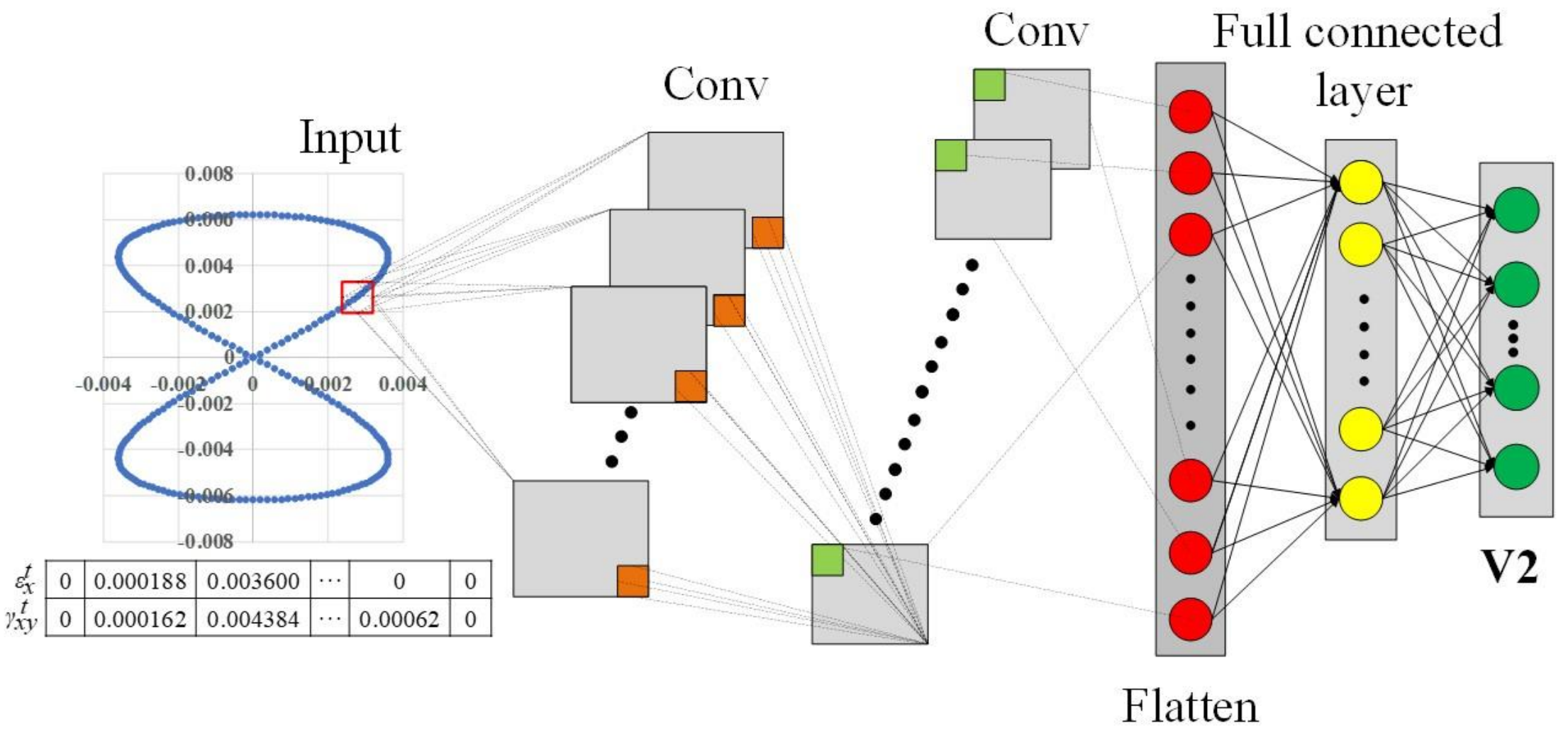

Fig. 9. CNN block feature extraction process

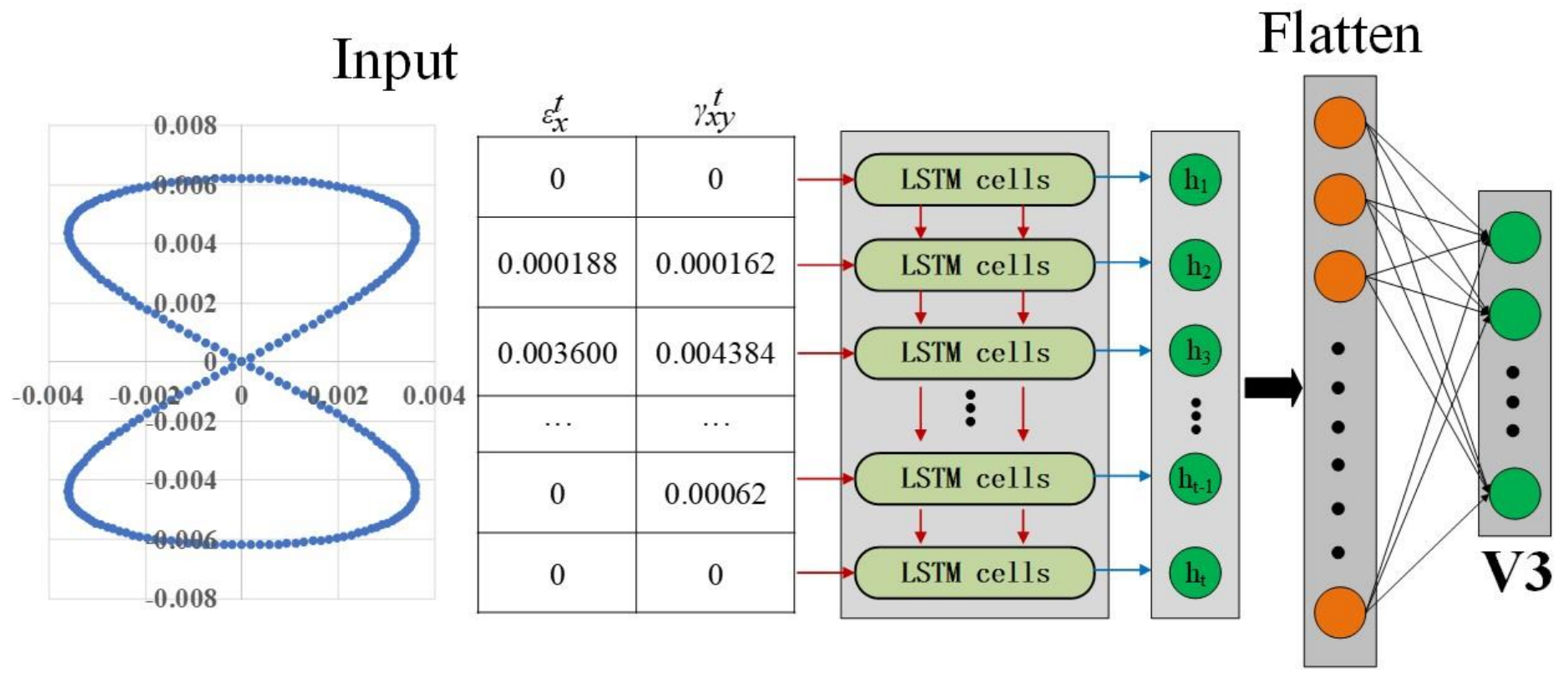

Fig. 10. LSTM block feature extraction process

The LSTM module employs a single layer of LSTM to extract features from the temporal information. The extracted features are then flattened into a one-dimensional vector and passed through a fully connected neural network layer to obtain the feature vector **V3**, with dimensions consistent with **V1** and **V2**. The principles of CNN[8, 9, 13, 41] and LSTM[6, 9] networks are detailed in the references cited, which are not the focus of this study. Fig. 10 illustrates how the LSTM neural network completes the feature extraction process for multiaxial loading path sequences.

In the frequency domain analysis module, the temporal loading information is transformed into frequency domain information for analysis. This transformation is achieved through Fourier transformation. In the proposed method, the input temporal data $X_{Li}$ is not directly subjected to Fourier transformation. Instead, an Fnet neural network is utilized. Fnet replaces the Attention layer in the Transformer[42] with a parameterless Fourier transformation, enabling the extraction of deep frequency domain features from the loading path information and their thorough integration with the original temporal data. The Discrete Fourier Transform (DFT) is defined as follows:

$$X_k = \sum_{n=0}^{N-1} X_{Li} e^{-\frac{2\pi i}{N} nk}, 0 \leqslant k \leqslant N-1 \tag{8}$$

*where* $X_{Li}$ represents the input sequence，t∈[0, N-1]. For each value of $k$, the Discrete Fourier Transform (DFT) computes a new representation $X_k$ as the summation of all the original input tokens $X_{Li}$. [20] The parameterless Fourier transformation used in Fnet is represented as:

$$y = \Re(F_{seq}(F_{hidden}(x))) \tag{9}$$

where $x$ represents the input time series, and $y$ denotes the transformed result. $\Re$ signifies taking the real part only, $F_{seq}$ denotes performing a Fourier transform along the sequence dimension, and $F_{hidden}$ represents performing a Fourier transform along the feature vector dimension. This equation first conducts a Fourier transform along the feature vector dimension, followed by another transformation along the sequence dimension, and finally takes the real part. This process is illustrated in the FNet block in Fig. 7.

Considering the duality of Fourier transformation, within the Fnet block, the F module, after stacking N blocks, continuously performs Fourier transformation and inverse transformation. The input information undergoes repeated transformations between the frequency domain and the time domain. Simultaneously, frequency domain features hidden within the data are also extracted. Finally, the frequency domain features are dimensionally transformed and input into a fully connected layer (FCY) to obtain the feature vector **V4**, with dimensions of batch size × 100.

### *3.2.2 Feature fusion and fatigue life prediction*

After the feature extraction, four feature vectors, **V1**, **V2**, **V3**, and **V4**, are obtained. These four two-dimensional vectors are concatenated into a three-dimensional feature vector **V** (Batch size × 4 × 100). Subsequently, through a self-attention layer, information from different feature extraction modules is integrated[7, 42, 43], and the information is weighted and fused to better capture the correlations and importance among multiple features.

Following the feature fusion, the final feature vector comprehensively incorporates the mechanical properties, geometric features, and deep features extracted through the integration of three perspectives of the loading path. This feature vector is then unfolded into a one-dimensional vector and fed into a three-layer fully connected neural network to establish the mapping relationship between the fused features and the fatigue life, resulting in the final predicted value.

### *3.3 Hyperparameters analysis and model parameters determination*

Hyperparameters are parameters that need to be manually set before training the model. They control the structure and learning process of the model, rather than being learned from the training data. The selection of hyperparameters has a significant impact on the performance and generalization ability of the model. The range of hyperparameters involved in the model is quite broad, including the number of network layers, the connectivity between layers, the number of convolutional kernels, and the size of convolutional kernels. Other hyperparameters, such as the learning rate, batch size, number of training epochs, type of optimizer, and loss function, are also crucial. Research on how to determine hyperparameters can be found in references[41, 44, 45]. This study does not focus on discussing this aspect. Some of the main hyperparameters used in this research are introduced in Table 2. In addition to the parameters of the neural network listed in the table, some optimization parameters also need to be determined.

The number of samples per input batch (Batch size) was set to 24, and the training epochs were determined to be 1000. Additionally, this study tested three types of activation functions: tanh, sigmoid, and ReLU. Since sigmoid and tanh are commonly used for classification tasks, and the model used in this study is suitable for regression tasks, the ReLU activation function (f(x)=max(0,x)) was selected due to its better training performance.[46, 47]Adam optimizer was chosen, with the learning rate set to 0.001.

The loss function evaluates the disparity between the predicted values and the actual values, with lower values indicating better model performance. Different models typically use different loss functions. This study compared various loss functions such as mean squared error (MSE), mean absolute error (MAE), and root mean square error (RMSE), among which MSE yielded the best results. MSE is one of the commonly used loss functions in regression problems, measuring the average of the squared distances between the model's predicted values $f(\mathrm{x})$ and the true values $y$ of the samples. Its formula is as follows:

$$\mathrm{MSE} = \frac{1}{m}\sum_{i=1}^{m}\left(y_i - \hat{y}_i\right)^2 \tag{10}$$

where $y_i$ and $\hat{y}$ represent the true value and predicted value, respectively, for the $i$th sample, with $m$ denoting the number of samples.

Table 2. Hyperparameters of the proposed multi-view neural network model

| Block | Item | Value |
|---|---|---|
| DNN block | Neurons of 1st FCY | 100 |
| | Neurons of 2nd FCY | 50 |
| | Neurons of 3rd FCY | 100 |
| CNN block | Kernel size of 1st convolution layer | 2×2 |
| | Kernel size of 2nd convolution layer | 1×1 |
| | FCY neurons of CNN block | 100 |
| LSTM block | Neurons of LSTM hidden layer | 5 |
| | $t$ (Sequence length of $\varepsilon_t$ and $\gamma_t$) | 241 |
| | FCY neurons of LSTM block | 100 |
| Fnet block | Number of F modules | 3 |
| | FCY neurons of Fnet block | 100 |
| Fatigue life prediction part | Neurons of 1st FCY | 100 |
| | Neurons of 2nd FCY | 200 |
| | Neurons of 3rd FCY | 100 |

## 4. Fatigue life prediction results and discussion

In this section, the predictive capabilities of the proposed model are discussed. Section 4.1 introduces the evaluation metrics used for the model assessment, while Section 4.2 randomly partitions the fatigue data of 19 materials into training and testing sets, conducting a comprehensive analysis of the predictive performance. To test the extrapolation ability of the model, Section 4.3 involves the partitioning of a specific test set containing complex loading paths for further analysis.

### *4.1 Model evaluation metrics*

To comprehensively evaluate the performance of the model, three evaluation metrics are adopted in this study: Mean Absolute Error (MAE), Mean Squared Error (MSE), and the Coefficient of Determination (R2). These metrics reflect the differences between the predicted values and the actual values[8, 48]. MAE measures the absolute value of the prediction error, while MSE measures the squared prediction error. Smaller values of MAE and MSE indicate better predictive performance of the model. The formula for calculating MAE is as follows:

$$\mathrm{MAE} = \frac{1}{m}\sum_{i=1}^{m}\left|y_i - \hat{y}\right| \tag{11}$$

$R^2$ is a metric that measures the goodness of fit of the observed values to the fitted regression line. It ranges from 0 to 1, with a value closer to 1 indicating a stronger applicability of the model. The formula for $R^2$ is:

$$R^2 = 1 - \frac{\sum_{i=1}^{n}\left(y_i - \hat{y}_i\right)^2}{\sum_{i=1}^{n}\left(y_i - \bar{y}_i\right)^2} \tag{12}$$

where $\bar{y}_i$ represents the mean of the true value of fatigue life, and the meanings of other symbols are consistent with those described in the above MSE.

### *4.2 Life prediction accuracy under multiaxial loading conditions*

After determining the model parameters, the model underwent training, during which the hyperparameters were continuously adjusted to identify the optimal model. The Mean Squared Error (MSE) of both the training and testing sets gradually decreased with the increase in training batches, as depicted in Fig. 11. Initially, at the onset of training, both the testing and training sets exhibited relatively high errors. However, as the number of training iterations increased, the MSE of both sets gradually converged to a lower level. Eventually, the training and testing errors became very close to each other, indicating a strong fitting effect of the model.

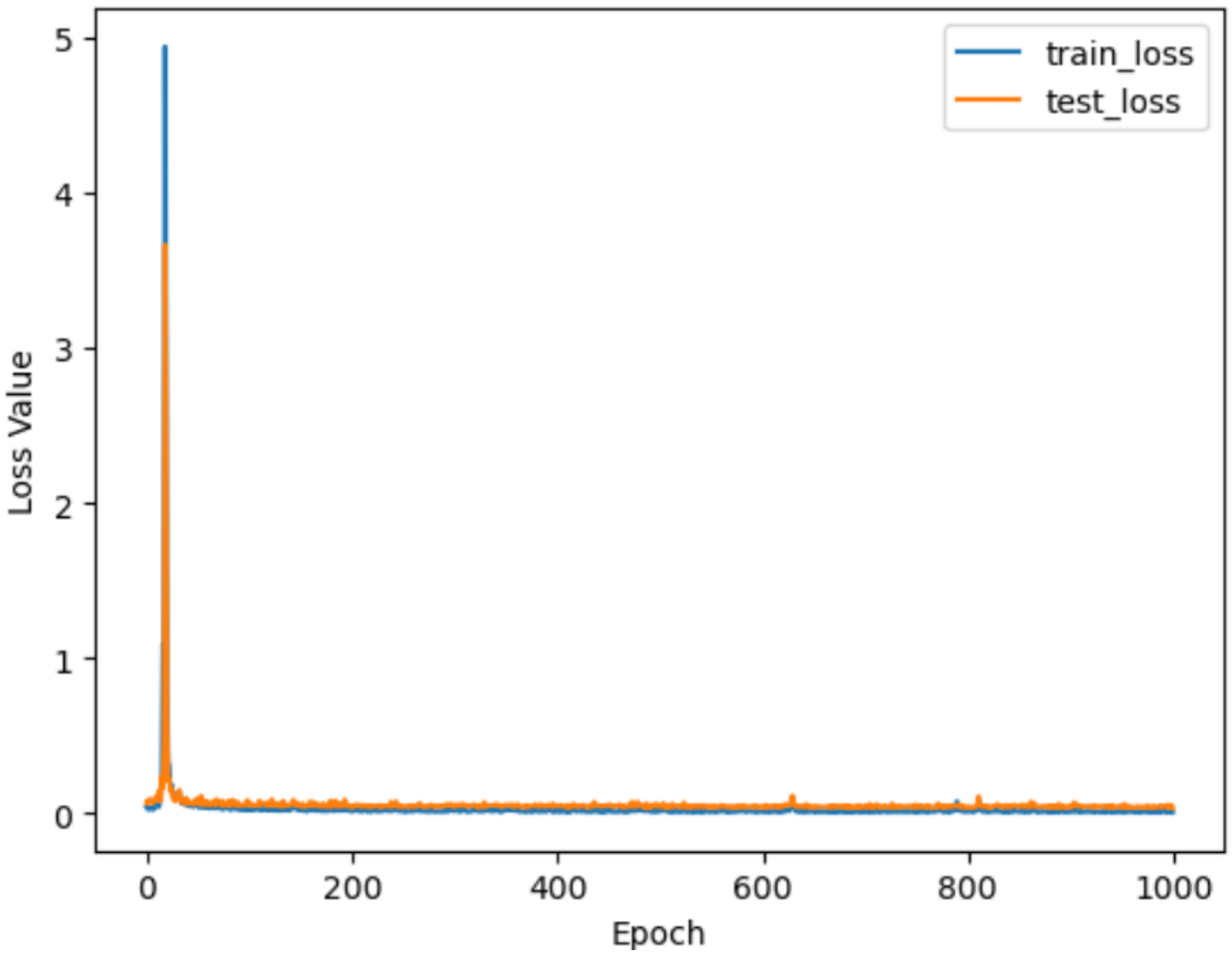

Fig.11 . Loss value of the training process.

The analysis of the model using the evaluation metrics introduced in Section 4.1 yielded the results presented in Table 2.

Table 3. Evaluation results of the proposed model

| Item | Range | Value |
|---|---|---|
| $R^2$ | [0,1] | 0.924856 |
| MSE | [0,+∞) | 0.130023 |
| MAE | [0,+∞) | 0.029773 |

It can be observed that the value of R2 has reached around 0.93, indicating a good fitting effect of the model. Both MSE and MAE values are less than 0.1, indicating high precision of the model.

The predictive results of the proposed model are illustrated in Fig. 12, where Fig. 12(a) and (b) depict the overall prediction performance on the training and test sets, respectively. Among the 445 samples in the training set, all samples are within 3 times the error dispersion band, with only 10 samples falling outside the 2 times dispersion band. Approximately 98% of the samples are within the 2 times dispersion band, and about 75% are within the 1.5 times dispersion band. The performance on the test set is similar to that of the training set, with most samples exhibiting good accuracy.

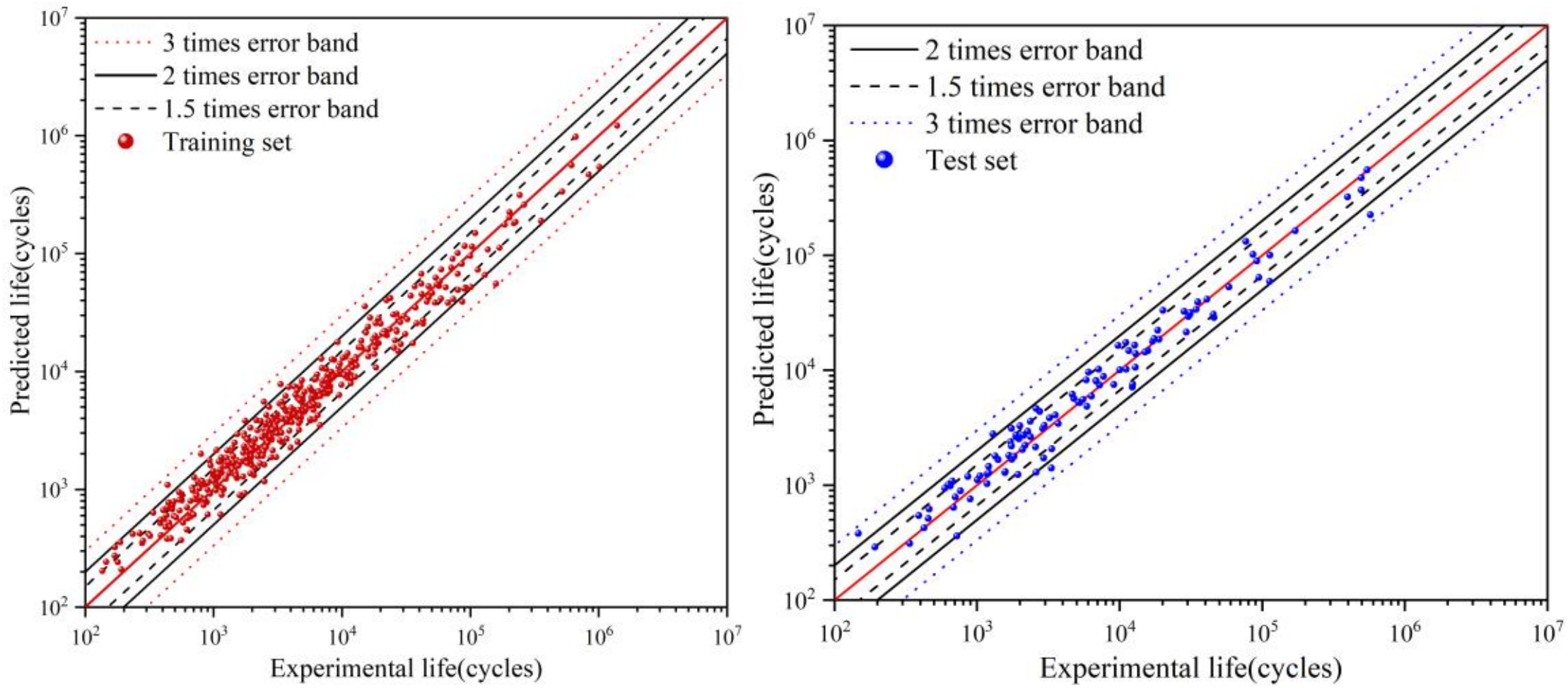

Fig. 12. Predictive results of the proposed model in (a) Training set; (b) Test set.

Due to space limitations, not all prediction results are presented. Instead, a few representative materials are selected for display and analysis. These materials' samples encompass a diverse range of loading paths, including uniaxial loading, proportional loading, sinusoidal non-proportional loading, trapezoidal non-proportional loading, triangular non-proportional loading, loading at different frequencies, and pseudo-random loading. Additionally, it is important to note that no additional training was conducted for these material predictions; the determined optimal model is a general model applicable to different materials.

Fig. 13 illustrates the prediction results for six different materials. The results indicate that the model can effectively predict the fatigue life under various loading paths for different materials. Samples under complex loading paths are mostly within the 2 times dispersion band, with only a few samples from materials E355 and PA38-T6 falling outside the 2 times dispersion band. One sample from material S460N under a 90° phase difference sinusoidal loading path with stress ratio 0 falls within the 2 times dispersion band, demonstrating excellent prediction accuracy. This indicates that: 1. The proposed deep learning method can effectively capture the deep features of fatigue loading paths; 2. The material-level features (mechanical properties, geometric features) selected for prediction can adequately represent the differences between different materials; 3. The deep learning model can effectively integrate material-level features with loading path-level features and establish the mapping relationship with the corresponding fatigue life. Even when predicting different materials with vastly different material characteristics and extremely complex loading paths, the proposed method demonstrates satisfactory prediction accuracy.

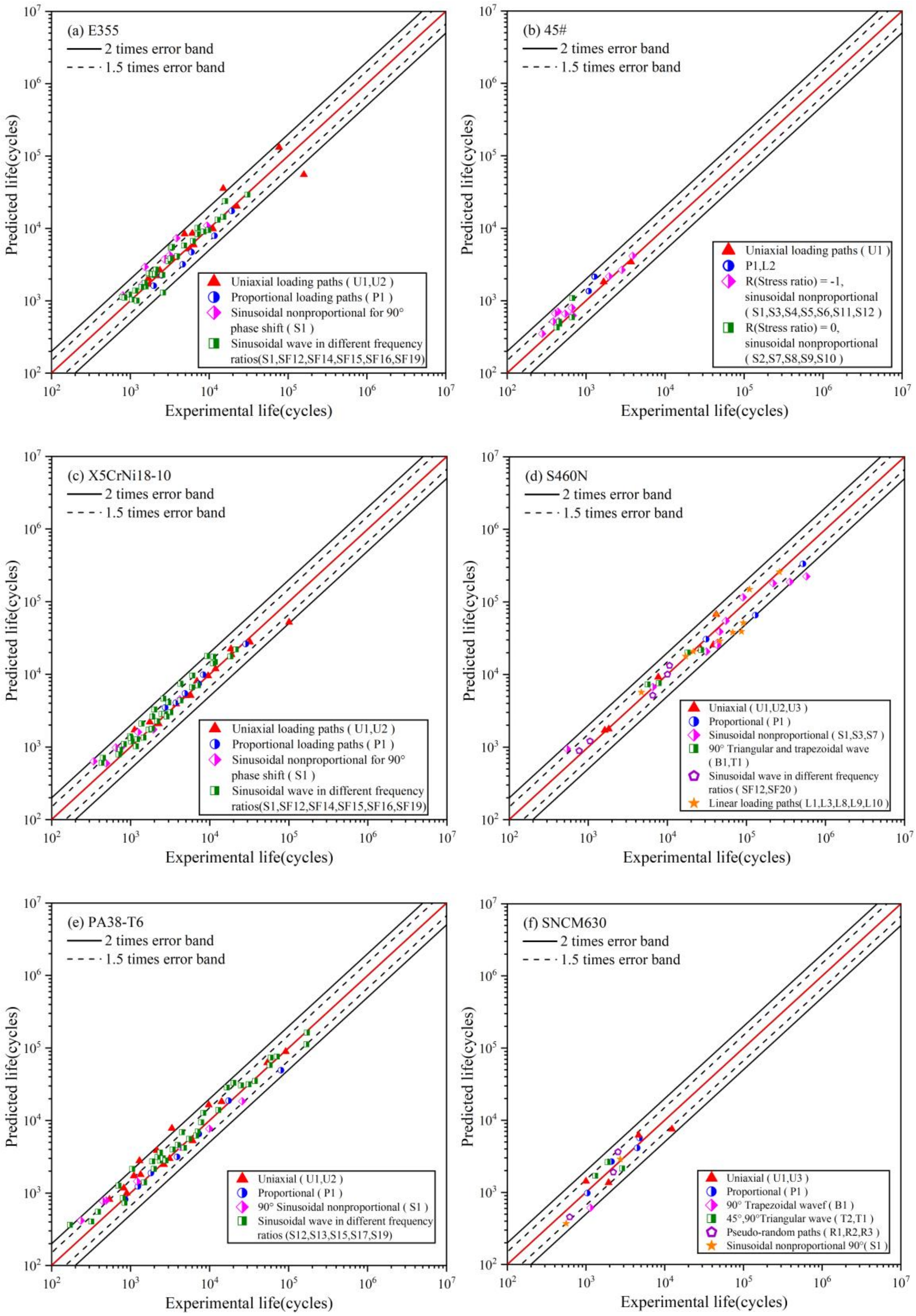

Fig. 13. Predicted results under different materials: (a) E355; (b) 45#; (c) X5CrNi18-10; (d) S460N; (e) PA38-T6; (f) SNCM630.

### *4.3 Extrapolation capability of the proposed method*

To assess the extrapolation capability of the model, the aforementioned 6 materials were used as separate test sets, and the model was retrained with the remaining materials as training sets. Specific predictive performance is illustrated in Fig. 14. Due to the variations in the test and training sets, some hyperparameter adjustments were necessary. During this process, the batch size was modified to 21, and the number of training epochs was set to 2000.

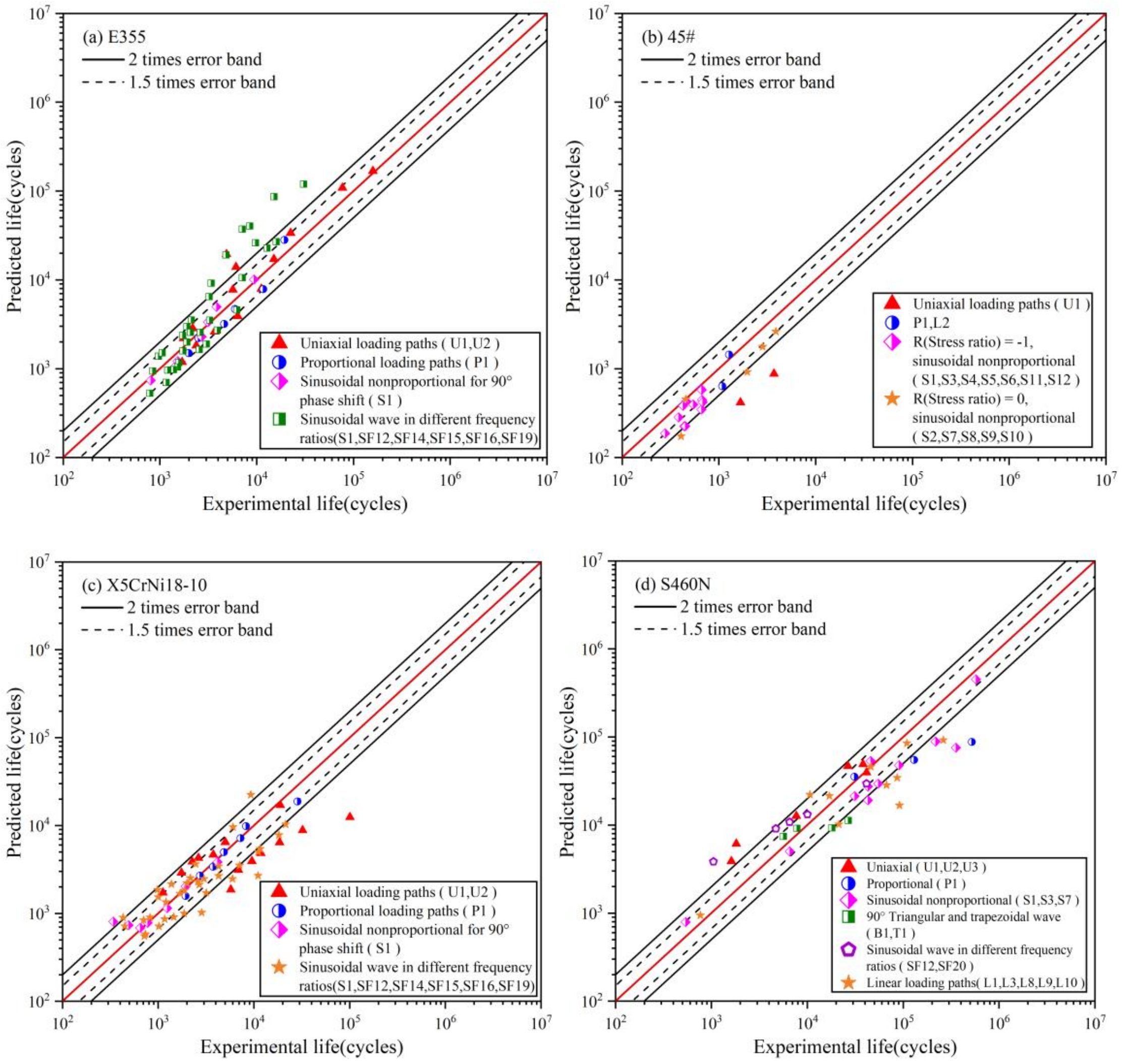

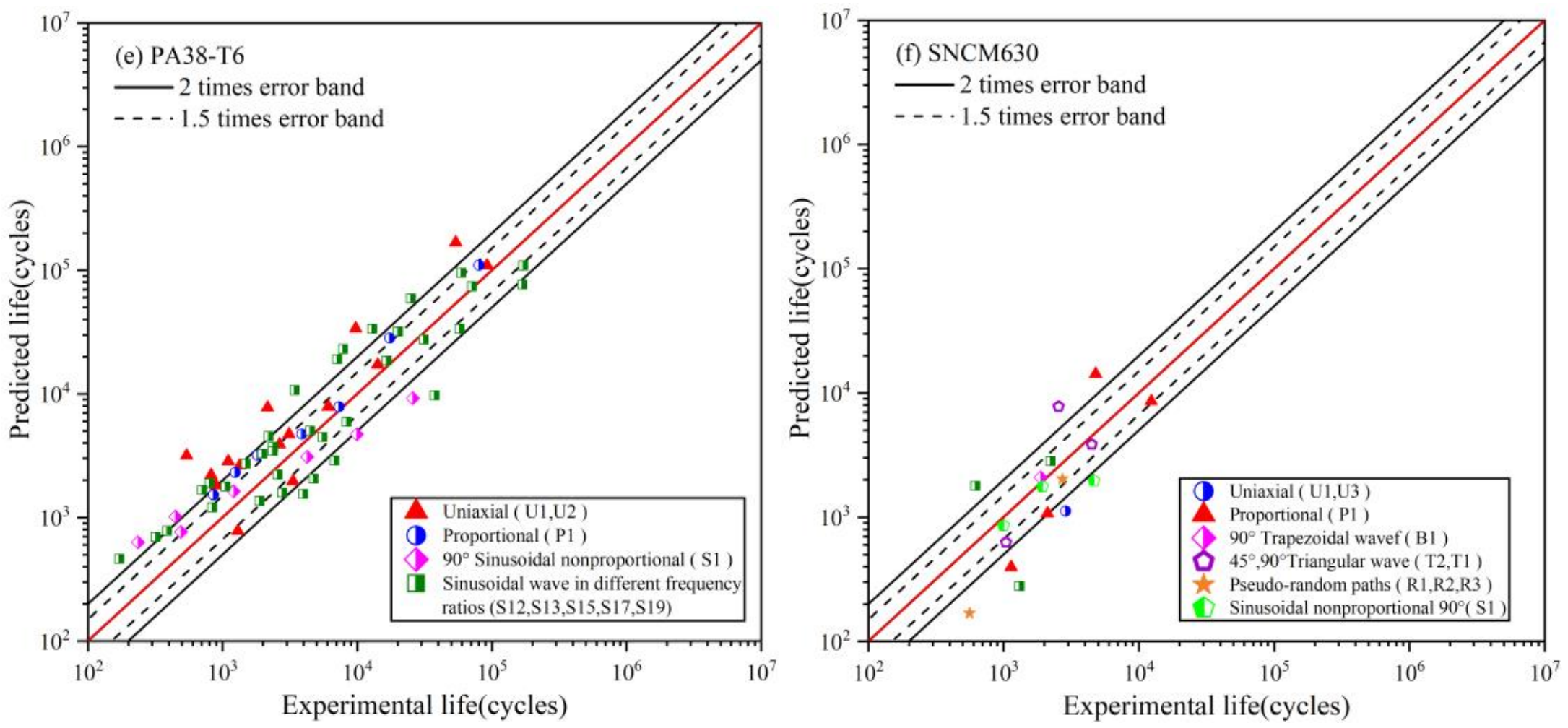

Fig. 14. The predictive performance of the model using six specific materials as separate test sets: (a) E355; (b) 45#; (c) X5CrNi18-10; (d) S460N; (e) PA38-T6; (f) SNCM630.

The predictive results indicate that the majority of predictions for these six materials fall within the twice scatter band, suggesting a good extrapolation capability of the model to unknown materials. However, there are noticeably more predictions outside the twice scatter band, with a few predictions deviating significantly from it.

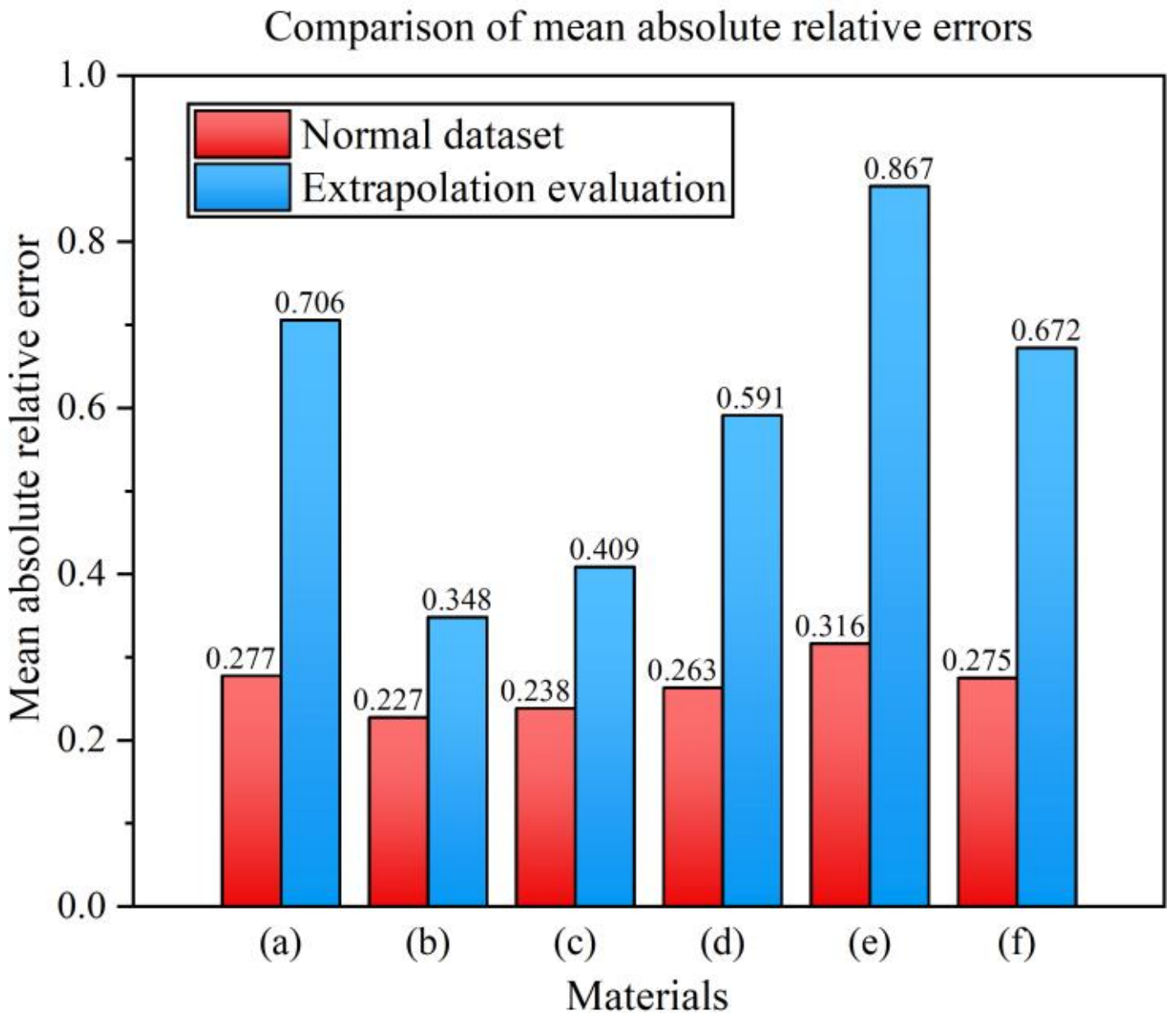

Fig.15. Comparison of mean relative errors for specific materials: (a)E355; (b) 45#; (c) X5CrNi18-10; (d) S460N; (e) PA38-T6; (f) SNCM630.

As depicted in Fig. 15, the mean absolute relative errors(Eq. (13)) of the predictions for the 6 materials have all increased: the relative errors for 45# and X5CrNi18-10 have increased less noticeably. For E355, S460N, PA38-T6, and SNCM630, the relative errors have approximately doubled. However, overall, the relative error values remain at a relatively low level, all below 1.0. This indicates that the model possesses strong feature extraction capabilities and can reasonably predict the fatigue life of unknown materials based on the data in the limited training set.

$$MARE = \frac{1}{m}\sum_{i=1}^{m}\left|\frac{\hat{y}_i - y_i}{y_i}\right| \tag{13}$$

Specifically, for the E355 material, most predictions outside the twice scatter band stem from loading paths with different frequency ratios of non-proportional loading. Predictive performance is notably poorer for the 45# and X5CrNi18-10 materials under uniaxial loading paths. PA38-T6 exhibits inferior performance under both uniaxial loading paths and loading paths with different frequency ratios of non-proportional loading. Conversely, for the relatively fewer samples of S460N and SNCM630, points with poor predictive performance do not solely originate from a single loading path; rather, they are relatively evenly distributed between uniaxial loading paths and complex multiaxial loading paths.

In summary, the model demonstrates the ability to predict fatigue life under different loading paths for various materials while maintaining a certain level of accuracy. This further indicates that the model can learn important factors affecting multiaxial fatigue life from the existing training set materials, thereby exhibiting good extrapolation capability. The possible reason for this could be the inconsistency in the distribution between the training and test sets. The training set may not adequately cover the entire sample space, which is a common issue in deep learning models. If the training set contains fewer samples or lacks diversity in sample types, it can lead to a decrease in accuracy in the test set. This issue can typically be addressed by constructing a larger and more diverse training set, encompassing a wider range of samples.

### *4.4 Discussion*

Currently, researchers mainly use LSTM and CNN for the analysis of multiaxial fatigue loading paths, focusing on a particular single perspective. This study attempts to introduce a frequency domain analysis neural network and integrate multi-perspective analysis methods for feature extraction from fatigue loading paths. The proposed approach achieves high accuracy in predicting 19 different materials, validating the effectiveness of the method. Furthermore, most existing neural network models are trained on samples from specific materials. While such models perform well in predicting the fatigue life of those specific materials, they require retraining when predicting other unknown materials, involving complex and tedious parameter tuning. This study addresses this issue by merging material-level features with loading path features to construct a neural network that can be applied to different materials simultaneously. The efficacy of this approach also underscores the ability of the selected material parameters to effectively represent variations among different materials. In future research, integrating external environmental features could be explored to investigate fatigue life under different types of failure modes.

Although a total of 557 samples were collected in this article, only 19 types of materials were included, which cannot cover the characteristics of all material types. In addition, the ideal dataset scenario should include a sufficient number and types of loading paths, and a sufficient number of samples corresponding to the fatigue life of each interval. The sample distribution shown in Fig. 6 of this study indicates that the number of samples in the range of ultra-low cycle fatigue and ultra-high cycle fatigue is relatively small, indicating that the dataset cannot truly represent the entire sample space. This leads to a decrease in the prediction accuracy of the model for unknown materials when verifying its extrapolation ability. Deep learning is a highly data-dependent method that requires fatigue data from a wider variety of materials to train the model. However, multiaxial fatigue tests incur high costs, rendering them ineffective as a primary data source. Therefore, considering the utilization of simulated fatigue data for training deep learning models, constructing a larger training dataset with a greater variety of materials and more complex loading paths may represent a promising new direction. Future research endeavors are likely to explore this possibility.

Additionally, the black-box nature of neural networks results in poor interpretability, as the extracted features are difficult to explain physically, limiting their application in fields with high interpretability requirements. To address this, introducing physics-based knowledge into neural networks is a promising direction. By incorporating physical equations as constraints into the neural network, the fitted results can better adhere to physical laws. Some studies have already made progress in using Physics-Informed Neural Networks (PINN) for multiaxial fatigue life prediction tasks[49–51].

## 5. Conclusions and future work

The study introduces frequency domain analysis and constructs a novel multiaxial fatigue life prediction model using integrated modeling, incorporating multi-perspective analysis. The performance of this model is validated using multiaxial fatigue experimental data from 19 different materials. Evaluation of the model is conducted using three metrics: $R^2$, MSE, and MAE. Additionally, the extrapolation capability of the model is explored on datasets composed of specific materials. The research findings are summarized as follows:

(1) The proposed method demonstrates good predictive performance for fatigue life across different materials, with the majority of predictions falling within 1.5 times the error dispersion band, and almost all data within 2 times the dispersion band.

(2) The material-level features selected effectively represent variations among different materials, while the proposed multi-perspective neural network adeptly extracts deep features from loading paths. Moreover, the self-attention mechanism within the model facilitates effective fusion of these two sets of features.

(3) Evaluation results based on the three metrics indicate favorable model fitting (as indicated by $R^2$ values) and low prediction errors (as indicated by MSE and MAE), further validating the predictive capability of the model.

(4) Despite the increased disparity between the distributions of training and testing sets, the model's predictive ability remains within an acceptable range of errors, indicating its strong extrapolation capability.

The model proposed in this study extracts features from both material and load perspectives. However, the data inputted into the model may not necessarily encompass all information related to these two aspects. Considering more comprehensive features may be a method to further improve prediction accuracy. Additionally, incorporating complex external environmental factors such as temperature and loading conditions as model features to address more complex fatigue failure problems (e.g., creep fatigue, high-temperature fatigue, corrosion fatigue) requires further research in the future.

## Acknowledgment

The authors would like to thank the support from the National Key R&D Program of China under Grant No. 2022YFB2602700, the National Natural Science Foundation of China (No. 52378216, and 52078080), and the National Natural Science Foundation of China for Excellent Young Scientists Fund (Overseas, No. HW2021006).